\pdfoutput=1
\documentclass[11pt,a4paper]{article}

\usepackage[utf8]{inputenc}
\usepackage[T1]{fontenc}
\usepackage{lmodern}
\usepackage[margin=2.2cm]{geometry}
\usepackage[round]{natbib}
\usepackage{amsmath,amssymb}
\usepackage{amsthm}
\usepackage{xcolor}
\usepackage{listings}
\usepackage{latexml}  
\iflatexml\else\usepackage[most]{tcolorbox}\fi
\usepackage{microtype}
\usepackage[scaled=0.95]{inconsolata}
\definecolor{leankw}{RGB}{0,51,153}
\usepackage[colorlinks, allcolors=leankw]{hyperref}

\newcommand{\leanlinkdef}[2]{\expandafter\def\csname leanz@#1\endcsname{#2}}
\leanlinkdef{ex1}{Tutorial}
\leanlinkdef{ex2}{Tutorial}
\leanlinkdef{Identification}{Identification}
\leanlinkdef{Tutorial}{Tutorial}
\leanlinkdef{Lemma_3_1_Ljung}{Lemma_3_1_Ljung}
\leanlinkdef{Template}{Template}
\edef\leanhash{\string#}
\begingroup\catcode`\%=12 \xdef\leanpct{
\newcommand{\leanraw}{https://raw.githubusercontent.com/manasdm/Lean-SysID/main/Lean\leanpct 20code}
\newcommand{\hrefx}[2]{\edef\hrefxurl{#1}\expandafter\href\expandafter{\hrefxurl}{#2}}
\newcommand{\leanlink}[2]{\hrefx{https://live.lean-lang.org/\leanhash url=\leanraw/\csname leanz@#1\endcsname.lean}{#2}}
\newcommand{\leanlive}[2]{\leanlink{#1}{\textcolor{leankw}{#2}}}
\newcommand{\mathlib}[3]{\hrefx{https://leanprover-community.github.io/mathlib4_docs/Mathlib/#1.html\leanhash #2}{\texttt{#3}}}
\newcommand{\leanbutton}[1]{\leanlink{#1}{{\setlength{\fboxsep}{1pt}\colorbox{leankw}{%
  \textcolor{white}{\sffamily\fontsize{5}{5}\selectfont$\blacktriangleright$\,Open in Lean Web}}}}}
\iflatexml\else
\tcbset{openlean/.style={enhanced, overlay={\node[anchor=north east, inner sep=0pt]
  at ([xshift=-1.5pt, yshift=-1.5pt]frame.north east) {\leanbutton{#1}};}}}
\fi

\DeclareUnicodeCharacter{03C1}{\ensuremath{\rho}}
\DeclareUnicodeCharacter{03B1}{\ensuremath{\alpha}}
\DeclareUnicodeCharacter{03B2}{\ensuremath{\beta}}
\DeclareUnicodeCharacter{03B4}{\ensuremath{\delta}}
\DeclareUnicodeCharacter{0394}{\ensuremath{\Delta}}
\DeclareUnicodeCharacter{03BC}{\ensuremath{\mu}}
\DeclareUnicodeCharacter{03A9}{\ensuremath{\Omega}}
\DeclareUnicodeCharacter{03C9}{\ensuremath{\omega}}
\DeclareUnicodeCharacter{03A3}{\ensuremath{\Sigma}}
\DeclareUnicodeCharacter{0398}{\ensuremath{\Theta}}
\DeclareUnicodeCharacter{03B8}{\ensuremath{\theta}}
\DeclareUnicodeCharacter{03A6}{\ensuremath{\Phi}}
\DeclareUnicodeCharacter{211D}{\ensuremath{\mathbb{R}}}
\DeclareUnicodeCharacter{2115}{\ensuremath{\mathbb{N}}}
\DeclareUnicodeCharacter{22A2}{\ensuremath{\vdash}}
\DeclareUnicodeCharacter{2192}{\ensuremath{\to}}
\DeclareUnicodeCharacter{2016}{\ensuremath{\|}}
\DeclareUnicodeCharacter{2264}{\ensuremath{\le}}
\DeclareUnicodeCharacter{2265}{\ensuremath{\ge}}
\DeclareUnicodeCharacter{2200}{\ensuremath{\forall}}
\DeclareUnicodeCharacter{2208}{\ensuremath{\in}}
\DeclareUnicodeCharacter{1D40}{\ensuremath{^{\top}}}
\DeclareUnicodeCharacter{1D65}{\ensuremath{_v}}
\DeclareUnicodeCharacter{2B1D}{\ensuremath{\cdot}}
\DeclareUnicodeCharacter{2297}{\ensuremath{\otimes}}
\DeclareUnicodeCharacter{2096}{\ensuremath{_k}}
\DeclareUnicodeCharacter{221A}{\ensuremath{\surd}}
\DeclareUnicodeCharacter{2026}{\ensuremath{\dots}}
\DeclareUnicodeCharacter{00B7}{\ensuremath{\cdot}}
\DeclareUnicodeCharacter{207B}{\ensuremath{^{-}}}
\DeclareUnicodeCharacter{00B9}{\ensuremath{^{1}}}
\DeclareUnicodeCharacter{2080}{\ensuremath{_0}}
\DeclareUnicodeCharacter{2203}{\ensuremath{\exists}}
\DeclareUnicodeCharacter{2081}{\ensuremath{_1}}
\DeclareUnicodeCharacter{2082}{\ensuremath{_2}}
\DeclareUnicodeCharacter{27E8}{\ensuremath{\langle}}
\DeclareUnicodeCharacter{27E9}{\ensuremath{\rangle}}

\newcommand{\R}{\mathbb{R}}
\newcommand{\calD}{\mathcal{D}}

\newtheorem{lemma}{Lemma}
\theoremstyle{definition}
\newtheorem{remark}{Remark}
\newtheorem{example}{Example}

\definecolor{leankw}{RGB}{0,51,153}
\definecolor{leantac}{RGB}{160,60,0}
\definecolor{leancmt}{RGB}{80,120,80}
\definecolor{leanbg}{RGB}{246,247,251}
\lstdefinelanguage{lean}{
  morekeywords=[1]{theorem,lemma,def,abbrev,example,structure,variable,namespace,end,
    section,open,import,noncomputable,fun,let,have,show,from,by,at,with,match,if,then,
    else,in,calc,Type,Prop,Set,scoped,where,instance},
  morekeywords=[2]{intro,intros,rintro,exact,refine,apply,rw,simp,simpa,linarith,
    nlinarith,positivity,gcongr,ring,field_simp,unfold,induction,cases,rcases,obtain,
    constructor,use,norm_num,change,sorry,trivial,decide,omega,ext,congr,rfl,rwa,set,
    trace_state,specialize,funext,ring_nf,simp_rw,abel,noncomm_ring},
  sensitive=true, morecomment=[l]{--}, morecomment=[s]{/-}{-/}, morestring=[b]",
}
\lstset{
  language=lean, basicstyle=\ttfamily, columns=fullflexible, keepspaces=true,
  keywordstyle=[1]\color{leankw}\bfseries, keywordstyle=[2]\color{leantac},
  commentstyle=\color{leancmt}\itshape, backgroundcolor=\color{leanbg},
  numbers=left, numberstyle=\scriptsize\color{gray}, numbersep=3pt, xleftmargin=9pt,
  frame=none, breaklines=true, breakatwhitespace=true, showstringspaces=false,
  tabsize=2, aboveskip=3pt, belowskip=3pt, inputencoding=utf8, extendedchars=true,
  upquote=true,
  literate=
      {ᵀ}{{$^{\top}$}}1 {ℝ}{{$\mathbb{R}$}}1 {ℤ}{{$\mathbb{Z}$}}1 {ŷ}{{$\hat{y}$}}1 {ℕ}{{$\mathbb{N}$}}1 {≤}{{$\le$}}1
    {≥}{{$\ge$}}1 {ρ}{{$\rho$}}1 {α}{{$\alpha$}}1 {β}{{$\beta$}}1 {δ}{{$\delta$}}1
    {Δ}{{$\Delta$}}1 {μ}{{$\mu$}}1 {Ω}{{$\Omega$}}1 {ω}{{$\omega$}}1 {Σ}{{$\Sigma$}}1
    {Θ}{{$\Theta$}}1 {ι}{{$\iota$}}1 {χ}{{$\chi$}}1 {κ}{{$\kappa$}}1 {λ}{{$\lambda$}}1
    {θ}{{$\theta$}}1 {Φ}{{$\Phi$}}1 {ε}{{$\varepsilon$}}1 {σ}{{$\sigma$}}1
    {φ}{{$\varphi$}}1 {γ}{{$\gamma$}}1 {τ}{{$\tau$}}1 {η}{{$\eta$}}1
    {‖}{{$\|$}}1 {→}{{$\to$}}1 {←}{{$\leftarrow$}}1 {↔}{{$\leftrightarrow$}}1
    {↦}{{$\mapsto$}}1 {⇒}{{$\Rightarrow$}}1 {⇔}{{$\Leftrightarrow$}}1
    {∀}{{$\forall$}}1 {∃}{{$\exists$}}1 {∈}{{$\in$}}1 {∉}{{$\notin$}}1
    {⊆}{{$\subseteq$}}1 {∧}{{$\wedge$}}1 {∨}{{$\vee$}}1 {¬}{{$\neg$}}1
    {√}{{$\surd$}}1 {∑}{{$\sum$}}1 {∘}{{$\circ$}}1 {×}{{$\times$}}1 {·}{{$\cdot$}}1
    {≠}{{$\neq$}}1 {⁻¹}{{$^{-1}$}}2 {²}{{$^2$}}1 {ⁿ}{{$^n$}}1 {⁺}{{$^+$}}1
    {₀}{{$_0$}}1 {₁}{{$_1$}}1 {₂}{{$_2$}}1 {ᵢ}{{$_i$}}1 {ⱼ}{{$_j$}}1 {ₖ}{{$_k$}}1
    {ᵥ}{{$_v$}}1 {⬝}{{$\cdot$}}1 {⊗}{{$\otimes$}}1 {⟨}{{$\langle$}}1 {⟩}{{$\rangle$}}1
    {⊢}{{$\vdash$}}1 {⪰}{{$\succeq$}}1 {∼}{{$\sim$}}1 {∝}{{$\propto$}}1
    {Â}{{$\hat{A}$}}1 {–}{{--}}1
    {ᾱ}{{$\bar\alpha$}}1 {•}{{$\bullet$}}1 {⊙}{{$\odot$}}1
    {θ̂}{{$\hat\theta$}}1 {∣}{{$|$}}1 {⋯}{{$\cdots$}}1 {…}{{$\dots$}}1
    {ₜ}{{$_t$}}1 {₃}{{$_3$}}1 {ᵃ}{{$^a$}}1 {⦃}{{$\{\!|$}}1 {⦄}{{$|\!\}$}}1
    {ᵐ}{{$^{m}$}}1 {∂}{{$\partial$}}1 {ℂ}{{$\mathbb{C}$}}1
}
\renewcommand{\topfraction}{0.9}
\renewcommand{\textfraction}{0.07}
\renewcommand{\floatpagefraction}{0.85}

\iflatexml
\renewcommand{\leanbutton}[1]{\leanlink{#1}{\colorbox{leankw}{\textcolor{white}{\textsf{\ ▶ Open in Lean Web\ }}}}}
\makeatletter
\newlength{\sbsL}\newlength{\sbsR}\setlength{\sbsL}{0.45\textwidth}\setlength{\sbsR}{0.5\textwidth}
\newcommand{\sbscols}{p{\sbsL}|p{\sbsR}}
\newcommand{\sbsitem}{\par\noindent\textbullet~}
\newenvironment{sbstab}{\par}{\par}
\newcommand{\sbsnext}{\par}
\newcommand{\sbsrow}[1][]{\par\medskip}
\def\lxb@label{}\def\lxb@open{}
\def\lxb@parse#1{\@for\lxb@kv:=#1\do{\expandafter\lxb@one\lxb@kv==\@nil}}
\def\lxb@one#1=#2=#3\@nil{\lxb@trim{#1}\lxb@key\lxb@trim{#2}\lxb@val
  \def\lxb@tmp{label}\ifx\lxb@key\lxb@tmp\let\lxb@label\lxb@val\fi
  \def\lxb@tmp{openlean}\ifx\lxb@key\lxb@tmp\let\lxb@open\lxb@val\fi}
\def\lxb@trim#1#2{\edef#2{\zap@space#1 \@empty}}
\newcommand{\lxb@button}{\ifx\lxb@open\@empty\else\leanbutton{\lxb@open}\fi}
\renewcommand{\figurename}{Box}  
\newenvironment{sbsrows}[2][]{\def\lxb@label{}\def\lxb@open{}\lxb@parse{#1}%
  \begin{figure}[h]\caption{#2}%
  \ifx\lxb@label\@empty\else\label{\lxb@label}\fi
  \noindent\lxb@button\par}{\end{figure}}
\newenvironment{leanbox}[1][]{\def\lxb@open{}\lxb@parse{#1}\par\noindent\lxb@button\par}{\par}
\newenvironment{goalbox}{\par}{\par}
\lstdefinestyle{goal}{numbers=none, backgroundcolor=\color{yellow!10},
  basicstyle=\ttfamily, commentstyle=\color{black}, keywordstyle=[1]\color{black},
  keywordstyle=[2]\color{black}}
\makeatother
\else
\newtcolorbox[auto counter]{sbs}[2][]{float=htbp, width=\textwidth, sidebyside,
  sidebyside align=top seam, lefthand width=0.385\textwidth, sidebyside gap=3mm,
  colback=white, colframe=black!45, boxrule=0.4pt, arc=1pt, left=2mm, right=1.5mm,
  top=1mm, bottom=1mm, fonttitle=\bfseries\fontsize{9}{11}\selectfont, coltitle=black, colbacktitle=black!7,
  title={Box~\thetcbcounter: #2}, segmentation style={black!30, dashed}, #1}
\newlength{\sbsL}\newlength{\sbsR}
\newcommand{\sbscols}{@{}p{\sbsL}@{\hspace{1.5mm}{\color{black!30}\vrule}\hspace{1.5mm}}p{\sbsR}@{}}
\newcommand{\sbsitem}{\par\noindent\hangindent=0.9em\hangafter=1\makebox[0.9em][l]{\textbullet}}
\newenvironment{sbstab}{\begin{tabular}{\sbscols}}{\end{tabular}}
\newcommand{\sbsnext}{&}
\newcommand{\sbsrow}{\\}
\newtcolorbox[use counter from=sbs]{sbsrows}[2][]{float=htbp, width=\textwidth,
  openlean/.style={enhanced, overlay={\node[anchor=east, inner sep=0pt]
    at ([xshift=-1.5pt]title.east) {\leanbutton{##1}};}},
  colback=white, colframe=black!45, boxrule=0.4pt, arc=1pt, left=2mm, right=1.5mm,
  top=1mm, bottom=1mm, fonttitle=\bfseries\fontsize{9}{11}\selectfont, coltitle=black, colbacktitle=black!7,
  title={Box~\thetcbcounter: #2},
  before upper={\setlength{\sbsL}{0.38\textwidth}%
    \setlength{\sbsR}{\dimexpr\linewidth-\sbsL-3mm-0.4pt\relax}%
    \lstset{basicstyle=\ttfamily\fontsize{7.5}{9}\selectfont,
      numberstyle=\tiny\color{gray}, aboveskip=-6.5pt, belowskip=-10pt}\small}, #1}
\lstdefinestyle{goal}{numbers=none, backgroundcolor=\color{yellow!10},
  basicstyle=\ttfamily, xleftmargin=2pt, aboveskip=1pt,
  belowskip=1pt, commentstyle=\color{black}, keywordstyle=[1]\color{black},
  keywordstyle=[2]\color{black}}
\newtcolorbox{goalbox}{colback=yellow!10, colframe=yellow!50!black!40, boxrule=0.3pt,
  arc=1pt, left=0.5mm, right=0.5mm, top=0.2mm, bottom=0.2mm}
\newtcolorbox{leanbox}[1][]{colback=leanbg, colframe=black!25, boxrule=0.3pt, arc=1pt,
  left=0.5mm, right=0.5mm, top=0.3mm, bottom=0.3mm,
  before upper={\lstset{basicstyle=\ttfamily,
    backgroundcolor={}}}, #1}
\fi

\makeatletter
\lst@AddToHook{TextStyle}{\def\lst@basicstyle{\ttfamily}\let\lst@ifbreaklines\iffalse}
\makeatother

\title{From Pen-and-Paper to Machine-Checked Proofs:\\ Lean for System Identification}

\author{Manas Mejari, Dario Piga, Marco Forgione\\[4pt]
  \small IDSIA Dalle Molle Institute for Artificial Intelligence, USI-SUPSI, Lugano, Switzerland}

\date{}

\begin{document}
\maketitle

\begin{abstract}
Mathematical proofs in scientific papers are often long and intricate, relying on chains of references to earlier results. This makes it difficult for readers and reviewers to verify each and every step.
Interactive theorem proving offers
a complement: the theorem and its proof are written as code, and every inference is checked by a small, trusted piece of software called the kernel. This paper
serves as an introduction to the Lean~4 proof assistant for a system
identification and control audience, assuming no prior exposure to formal methods.
We first show how a Lean proof is read using two elementary examples: the first on identifiability, the second on set invariance.
We then illustrate how Lean can be used to analyse results from the literature,
taking as example a classic lemma on the inversion of the noise filter
in prediction-error identification \citep[Lemma~3.1]{ljung1999sysid}. Since in Lean every step must be justified, the
formalization brings to light a subtle point: the original claim of that lemma holds \emph{almost surely},
although not for every realisation, as its statement may suggest. A simple
counterexample makes this concrete.
Finally, the paper discusses guidelines and perspectives on the use of Lean in system identification research.

\end{abstract}

\section{Introduction}

Theoretical results in system identification and control often come with intricate proofs built on, \emph{e.g.},
pages of matrix inequalities, probabilistic arguments, and asymptotic analyses.
Delicate steps involving limits, expectations, or modes of
convergence are sometimes condensed or omitted, often for readability, for lack of space, to suit
a mixed audience of engineers and mathematicians, or simply by oversight. Refereeing such
proofs in detail is demanding, thus, implicit assumptions, gaps, imprecisions, and occasionally
outright errors still reach the published literature.

\emph{Interactive theorem provers} (proof assistants) offer a
complement. The authors write the proof in a formal language, and a small, trusted
program, the so-called \emph{kernel}, checks every step. Landmark results in mathematics have been
verified this way, including the Kepler conjecture, implemented in HOL Light and Isabelle
\citep{hales2017kepler}, and, more recently, the polynomial Freiman--Ruzsa conjecture
\citep{gowers2025marton}, whose proof was formalized in Lean within weeks of its
announcement. In September 2026, OpenAI announced an AI-generated proof, formalized in Lean, of finite-time blow-up for a forced
three-dimensional incompressible Navier--Stokes system~\citep{openai2026navierstokes}, at the time of writing of this paper, the result
has not been independently reviewed.

These tools have also been explored  in the control community. Examples include proofs, in Coq (now Rocq), of a
swing-up controller for the inverted pendulum \citep{rouhling2018pendulum} and of
LaSalle's invariance principle \citep{cohen2017lasalle},
a tutorial-style Lyapunov stability proof in \emph{KeYmaera X} aimed at control
engineers \citep{gleirscher2024stability}, a  Lean library for stability
and small-gain theory \citep{doll2026controllean} and
 recent  \emph{Lean for Control}~\citep{gokhale2026} and \emph{Control Theory in Lean}~\citep{forgione2026} repositories containing
control theoretic proofs in Lean.
Despite these efforts, proof assistants are not yet widely adopted in the control community.
In system identification, to the best of our knowledge, no such effort has been reported to date.

This paper serves as an introduction to Lean~4 \citep{deMoura2021lean4} and Mathlib
\citep{avigad2025mil}, its library of standard mathematical definitions and theorems on
which new proofs can be built conveniently. It is intended for readers with a background in system
identification and control, without assuming any prior exposure to proof assistants or formal verification methods.

Our contributions are threefold. First, Section~\ref{sec:lean} walks the reader
through a warm-up proof and two elementary examples, on identifiability under
persistent excitation and on the invariance of an interval for a linear system,
each shown side by side with its pen-and-paper proof. Second,
Section~\ref{sec:inv} illustrates the use of Lean to analyse published
results, taking as example \citet[Lemma~3.1]{ljung1999sysid} on the inversion of
the noise filter. As is common in textbooks, Ljung's proof condenses some routine
steps, however, formalizing it in Lean requires each of them to be justified. One step in particular,
the interchange of an infinite double sum, needs a technical condition that is not automatic,
and we provide a counterexample showing that it cannot be dropped. The condition
is satisfied for bounded noise, for which the original claim holds unchanged.
For white noise with finite variance, the standing assumption of the book, the
condition, and hence the claim, holds \emph{almost surely}, not for all realisations; the lemma
omits this detail.
Third, Section~\ref{sec:template} proposes a four-step template and guidelines for
writing Lean code in system identification, illustrated on the identifiability of
ARX models.

Writing and reading Lean proofs requires practice, and AI agents built on large
language models (LLMs) increasingly assist with both~\citep{yang2023leandojo}. The
agent itself need not be trusted: whoever, or whatever, wrote a proof, Lean accepts
it only after its kernel checks every step. Consequently, the user only needs to trust this
kernel, which is widely scrutinised and has independent implementations, and to make sure that
the statement formalized in Lean faithfully expresses the intended claim (see Section~\ref{sec:discussion}).

The paper is accompanied by Lean code\footnote{\url{https://github.com/manasdm/Lean-SysID}, tested with Lean v4.35.0.} that compiles with Lean~4 and the matching Mathlib release.
Every theorem, including the Mathlib results it builds on, relies only on the standard axioms of
classical mathematics built into Lean's logic, with no unproved steps and no additional axioms. All hypotheses are stated
explicitly in each theorem (see Section~\ref{sec:lean}).

\section{Lean in a nutshell}\label{sec:lean}

\subsection{What is Lean?}

Lean~4 is an interactive theorem prover based on dependent type
theory, in which statements and proofs are written in a functional
programming language and checked by the computer. Three ideas are central: (i)~statements are types and proofs are terms: just as a real number $\rho$ has type $\mathbb{R}$
(written \texttt{ρ : ℝ}), a proof \texttt{h} of $\rho<1$ has type $\rho<1$ (written \texttt{h : ρ < 1}); checking a proof thus amounts to checking that a term has the claimed type,
a mechanical task performed by the kernel; (ii)~proofs are usually written in \emph{tactic mode}: as a sequence of commands (\emph{tactics}) that transform the current \emph{goal},
initially the statement itself and then whatever remains to be proved, until Lean reports \texttt{No goals}, which means that the original statement is proven; (iii)~proofs typically build on established results: instead of re-deriving
known facts, one invokes theorems already proved, for instance in Mathlib, on linear algebra, analysis, or
probability, much as one would cite a textbook.

\subsection{A first example}

Suppose we want to prove the following trivial fact:
\begin{lemma}[example]\label{lem:example}
 Let $\rho \in \R$ satisfy the assumption $\rho < 1$. Then $0 < 1- \rho$.
\end{lemma}
\begin{proof}
By simple algebraic manipulation, we have $\rho < 1  \Rightarrow \rho-1 <0 \Rightarrow 0 <1 -\rho $.
\end{proof}

\emph{Lean proof:} A Lean proof starts with the claim to be proved (the current \texttt{goal}), which is progressively reduced to simpler
goals until each can be closed by a known fact or a hypothesis.
In Lean, the statement of Lemma~\ref{lem:example} and its proof are written as
\begin{leanbox}[openlean=ex1]
\begin{lstlisting}
import Mathlib
example (ρ : ℝ) (h : ρ < 1) : 0 < 1 - ρ := by
  linarith
\end{lstlisting}
\end{leanbox}
\noindent The code mirrors Lemma~\ref{lem:example}, piece by piece. After \texttt{example} (an unnamed theorem) are the
arguments: a real number \texttt{ρ} and a hypothesis \texttt{h} stating $\rho<1$; after the colon
comes the statement to be proved, namely $0<1-\rho$; finally, \texttt{:= by} opens the proof in tactic
mode, here a single tactic is written on line~3.
The editor (VS Code, or the browser at \leanlive{ex1}{\nolinkurl{https://live.lean-lang.org}}, opened with
its code by the button \leanbutton{ex1} on each listing) shows the \emph{goal state} next to the
cursor. With the cursor at the end of line~2, the tactic state on the right panel shows the following:
\begin{goalbox}
\begin{lstlisting}[style=goal]
Tactic state
1 goal
ρ : ℝ
h : ρ < 1
⊢ 0 < 1 - ρ
\end{lstlisting}
\end{goalbox}
Here, the \textit{Tactic state} shows \texttt{1 goal}. A goal consists of
\emph{assumptions} and a \emph{conclusion}, separated by the turnstile symbol
\texttt{⊢}. The assumptions, listed above \texttt{⊢}, are the objects and facts the proof may
currently use: arbitrary elements of some type, such as the
real number \texttt{ρ}, statements taken as true, such as the hypothesis \texttt{h : ρ < 1},
and intermediate results proved earlier. The conclusion, after \texttt{⊢}, is the statement that remains to
be proved. The tactic \texttt{linarith} (short for \emph{linear arithmetic}, line~3) is a Mathlib
automation that proves goals following from linear equalities and inequalities among the
assumptions; it closes this goal in a single step, and the tactic state becomes
\texttt{No goals}.

The same lemma can also be proved without \texttt{linarith}, using two basic mechanisms of Lean: (i) rewriting the
goal with a known equivalence, (ii) closing it with an assumption.

\begin{leanbox}[openlean=ex2]
\begin{lstlisting}
import Mathlib
example (ρ : ℝ) (h : ρ < 1) : 0 < 1 - ρ := by
    rw [sub_pos]   -- rewrite the goal  0 < 1 - ρ  into  ρ < 1
    exact h        -- this is exactly the hypothesis h: no goals left
\end{lstlisting}
\end{leanbox}
Here, line 3 uses the Mathlib lemma \mathlib{Algebra/Order/Group/Unbundled/Basic}{sub_pos}{sub\_pos}, which is a proof of the equivalence
\begin{equation}
\label{eq:sub_pos}
0 < a - b \iff b < a, \qquad \forall a,b \in \mathbb{R}.
\end{equation}
The  \texttt{rw} (rewrite) tactic uses the equivalence \texttt{sub\_pos} in ~\eqref{eq:sub_pos} from left to right
to rewrite the goal $0 < 1 - \rho$ into $\rho < 1$. We do not need to
specify $a$ and $b$, Lean matches the left-hand side of~\eqref{eq:sub_pos}
against the goal and finds $a = 1$ and $b = \rho$.
This turns the goal exactly into hypothesis \texttt{h} ($\rho <1$) as specified explicitly in line 4, after which the tactic state becomes \texttt{No goals} and the theorem is proved.

\begin{remark}
    A way to reach \texttt{No goals} without providing a full proof is the placeholder \texttt{sorry}, which closes any goal. Lean then
warns \texttt{declaration uses `sorry'} and \texttt{\#print axioms} lists
\texttt{sorryAx} among the axioms the theorem depends on. For all theorems in this paper, \texttt{\#print axioms} reports only
Lean's three standard axioms (\texttt{propext}, \texttt{Classical.choice},
\texttt{Quot.sound}). This is the expected set of axioms for a sound proof.
It indicates that there are no incomplete steps or custom, user-defined axioms.
\end{remark}

We now illustrate, through two toy examples from system identification and control, how results in these fields can be formalized in Lean.
A table of the Lean notation and tactics used in this paper is available in the accompanying GitHub repository.

\subsection{Identifiability and persistent excitation}

Parametric identification aims to recover unknown system parameters from observed input-output data. This relies on \emph{persistent excitation} (PE), \emph{i.e.},
the input must be sufficiently rich. Without sufficient excitation, the data do not determine the parameters uniquely, and the system is unidentifiable.
We formalize these concepts in the following lemma for a scalar system.

\begin{lemma}\label{lem:toyid}
Consider a scalar data-generating system $y_k =  \varphi_k \theta$, where $\theta \in  \mathbb{R}$ represents the unknown true system parameter, $\varphi_k \in \mathbb{R}$ is   the  regressor, $y_k \in \mathbb{R}$ is the measured  output at time  $k$. Let $\mathcal{D} \triangleq \{\varphi_1, y_1\}$ be the data collected at a single step from this system, and  let $\hat{\theta} \triangleq y_1 / \varphi_1$ be the single-step algebraic estimator for the parameter $\theta$. Then, the
true parameter $\theta$ is identifiable from $\mathcal{D}$, \emph{i.e.},
$$ \varphi_1 \theta_1 = \varphi_1 \theta_2 \Rightarrow \theta_1 = \theta_2, \forall \theta_1, \theta_2 \in \R,  $$
if and only if the regressor satisfies the single-step persistent excitation (PE) condition
\begin{equation}
    \varphi_1 \neq 0,
\end{equation}
under which the estimator is exact, $\hat{\theta} = \theta$.
\end{lemma}
\begin{proof} We prove both directions of the equivalence.

 (\emph{Only if}) Identifiability implies \(\varphi_1 \neq 0\): Suppose that \(\theta\) is identifiable from \(\mathcal{D}\), \emph{i.e.},
$
\varphi_1\theta_1=\varphi_1\theta_2
\implies \theta_1=\theta_2
$
for all \(\theta_1,\theta_2\in\mathbb{R}\). If \(\varphi_1=0\), then
$
\varphi_1\theta_1=\varphi_1\theta_2=0
$
for any \(\theta_1,\theta_2\in\mathbb{R}\), including \(\theta_1\neq\theta_2\), contradicting identifiability. Hence, \(\varphi_1\neq 0\).

(\emph{If}) \(\varphi_1\neq 0\) implies identifiability: Suppose that \(\varphi_1\neq 0\) and $\varphi_1\theta_1=\varphi_1\theta_2.$ Dividing both sides by \(\varphi_1\) yields
$\theta_1=\theta_2.$ Therefore, \(\theta\) is identifiable from \(\mathcal{D}\).

Finally, since the data are generated as \(y_1=\varphi_1\theta\), whenever \(\varphi_1\neq 0\), the estimator satisfies
$
\hat{\theta}=\frac{y_1}{\varphi_1}
=\frac{\varphi_1\theta}{\varphi_1}
=\theta.
$
Thus, the estimator recovers the true parameter.
\end{proof}

\begin{sbsrows}[label=box:toyid, openlean=Identification]{Identifiability from one step: Lemma~\ref{lem:toyid} (left)
and \leanlive{Identification}{\texttt{Identification.lean}} (right)}
\begin{sbstab}
\textbf{Definition} (L1--3). $\theta$ is identifiable from $\calD$ if different
parameters give different data: $\varphi_1\theta_1=\varphi_1\theta_2$ implies
$\theta_1=\theta_2$. The \texttt{def} gives this proposition (a \texttt{Prop}) the name
\lstinline!Identifiable φ₁!.
\sbsnext
\begin{lstlisting}
-- θ is identifiable from D = {φ₁, y₁}: equal outputs force equal θ.
def Identifiable (φ₁ : ℝ) : Prop :=
  ∀ θ₁ θ₂ : ℝ, φ₁ * θ₁ = φ₁ * θ₂ → θ₁ = θ₂
\end{lstlisting}
\sbsrow[4pt]
\textbf{Identifiability} (L5--6): \lstinline!Identifiable φ₁ ↔ φ₁ ≠ 0!.

\smallskip
\textbf{Proof.} \texttt{constructor} (L7) splits the equivalence into its two implications,
one bullet \texttt{·} each (L8, L11).
\sbsnext
\begin{lstlisting}[firstnumber=5]
-- Lemma 1: θ is identifiable from D if and only if φ₁ ≠ 0 (PE).
theorem identifiable_iff_PE (φ₁ : ℝ) : Identifiable φ₁ ↔ φ₁ ≠ 0 := by
  constructor
\end{lstlisting}
\sbsrow[2pt]
\sbsitem \emph{Only if} (L8--10): since $\varphi_1\neq0$ means ``$\varphi_1=0$ implies
\texttt{False}'', \texttt{intro} adds to the assumptions both \lstinline!hid : Identifiable φ₁!
and \lstinline!hφ : φ₁ = 0!, and the goal becomes \lstinline!⊢ False!. With $\varphi_1=0$,
$\theta_1=0$ and $\theta_2=1$ give the same output (checked by \texttt{simp}): zero
excitation cannot tell them apart. So identifiability yields \lstinline!h01 : 0 = 1!
(L9), which \texttt{norm\_num} refutes (L10).
\sbsnext
\begin{lstlisting}[firstnumber=8]
  · intro hid hφ                      -- only if: suppose φ₁ = 0
    have h01 := hid 0 1 (by simp [hφ]) -- θ = 0, θ = 1: same output
    norm_num at h01                   -- h01 : 0 = 1, contradiction
\end{lstlisting}
\sbsrow[2pt]
\sbsitem \emph{If} (L11--12): \texttt{intro} adds \lstinline!hφ : φ₁ ≠ 0!, $\theta_1$, $\theta_2$
and \lstinline!hy : φ₁ * θ₁ = φ₁ * θ₂! to the assumptions; cancelling $\varphi_1$
(Mathlib lemma \texttt{mul\_left\_cancel₀}) gives $\theta_1=\theta_2$.
\sbsnext
\begin{lstlisting}[firstnumber=11]
  · intro hφ θ₁ θ₂ hy              -- if: φ₁ ≠ 0 and φ₁ θ₁ = φ₁ θ₂
    exact mul_left_cancel₀ hφ hy   -- cancel φ₁
\end{lstlisting}
\sbsrow[4pt]
\textbf{Estimator} (L14--16). If $y_1=\varphi_1\theta$ and $\varphi_1\neq0$, then
$\hat\theta=y_1/\varphi_1=\theta$.
\sbsnext
\begin{lstlisting}[firstnumber=14]
-- Lemma 1 (estimator): under PE the estimate θ̂ = y₁ / φ₁ is the true θ.
theorem estimator_exact (θ φ₁ y₁ : ℝ) (hdata : y₁ = φ₁ * θ)
    (hφ : φ₁ ≠ 0) : y₁ / φ₁ = θ := by
\end{lstlisting}
\sbsrow[2pt]
\textbf{Proof.} Substitute the data (L17); the goal becomes
\lstinline!⊢ φ₁ * θ / φ₁ = θ!. Cancel $\varphi_1$ (L18). The tactic
\texttt{field\_simp} simplifies and clears denominators, and it may cancel $\varphi_1$ only because
the hypothesis \lstinline!hφ : φ₁ ≠ 0! is in the context.
\sbsnext
\begin{lstlisting}[firstnumber=17]
  rw [hdata]        -- goal becomes  φ₁ * θ / φ₁ = θ
  field_simp        -- cancel φ₁; allowed because hφ : φ₁ ≠ 0
\end{lstlisting}
\end{sbstab}
\end{sbsrows}
\emph{Lean proof:}
Box~\ref{box:toyid} proves Lemma~\ref{lem:toyid} in Lean, together with an explanation in the left panel. We first state the identifiability condition as a proposition in \texttt{Identifiable}.  Then, the theorem \texttt{identifiable\_iff\_PE} proves the necessary and sufficient condition that $\theta$ is identifiable if and only if $\varphi_1 \neq 0$ stated as \lstinline!Identifiable φ₁! \lstinline!↔ φ₁ ≠ 0!. The theorem is proved with a \texttt{constructor} tactic, which splits the equivalence into its two directions.

Next, in \texttt{estimator\_exact}, it is proved that the algebraic estimator is the true parameter $\hat{\theta} = \theta = \frac{y_1}{\varphi_1}$, under the PE condition.
Note that Lean needs \lstinline!hφ! (\emph{i.e.}, PE condition $\varphi_1 \neq 0$) explicitly as a hypothesis in \texttt{estimator\_exact} to prove that $\theta = \frac{y_1}{\varphi_1}$.
In Lean, division is
defined for every real number, with $y/0=0$ (the Mathlib lemma \mathlib{Algebra/GroupWithZero/Basic}{div_zero}{div\_zero}).
Without the hypothesis \lstinline!hφ! the claim would be false, since under zero excitation the
``estimate'' $y_1/\varphi_1=0/0$ equals $0$, whatever $\theta$ is. Thus, Lean forces us to
state the PE condition, and the ``only if'' part of Lemma~\ref{lem:toyid} shows that it
cannot be dropped.

\subsection{Invariant set of a linear dynamical system}
\label{sec:IS}
Invariant sets are powerful tools for analysing safety in control systems~\citep{blanchini1999}. A set is invariant for a system if every trajectory that starts in the set remains in it for all future time.
Next, we state and prove the following lemma, which gives a sufficient condition for the invariance of an interval.

\begin{lemma}\label{prop:IS}
For a scalar discrete-time system $x_{k+1}=ax_k$, if $|a|\le1$ then any interval set $\mathcal{R}:= [-r,r]$ is invariant, \emph{i.e.}, $x_0 \in \mathcal{R} \Rightarrow x_{k} \in \mathcal{R}$ for all $k \in \mathbb{N}$.
\end{lemma}
\begin{proof}
 We prove the result by mathematical induction.
Base case ($k=0$): By assumption, $x_0\in\mathcal{R}$. Hence, $|x_0|\leq r.$ Inductive step: Suppose that $x_k\in\mathcal{R}$ for some $k\in\mathbb{N}$, \emph{i.e.}, $|x_k|\leq r$. From the system dynamics $x_{k+1}=ax_k$, we obtain
$|x_{k+1}|=|ax_k|=|a||x_k|$. Since $|a|\leq 1$, it follows that
$|x_{k+1}|\leq |x_k|\leq r.$ Thus, $x_{k+1}\in\mathcal{R}$. By mathematical induction, $x_k\in\mathcal{R}$ for all $k\in\mathbb{N}$. Hence, $\mathcal{R}$ is invariant.
\end{proof}

\emph{Lean proof:}
In Box~\ref{box:scalar}, we provide the proof of Lemma~\ref{prop:IS}  with Lean code. In particular, we first prove the one-step statement (Claim 1) as a chain of
inequalities, written with Lean's \texttt{calc} construct, which mirrors the
theoretical proof line by line, and then the invariance condition for all $k \in \mathbb{N}$ (Claim 2), proved by induction
on $k$, as in the definition of an invariant set. The proof ends in the state \texttt{No goals}, proving Lemma~\ref{prop:IS}.

\begin{sbsrows}[label=box:scalar, openlean=Tutorial]{Invariance of an interval: Lemma~\ref{prop:IS} (left) and \leanlive{Tutorial}{\texttt{Tutorial.lean}} (right)}
\begin{sbstab}
\textbf{Claim 1.} If $|a|\le1$ and $|x|\le r$, then $|ax|\le r$ (L2--3).
\sbsnext
\begin{lstlisting}
-- The interval {x | |x| ≤ r} is invariant for x⁺ = a x when |a| ≤ 1.
theorem interval_invariant (a r x : ℝ) (ha : |a| ≤ 1) (hx : |x| ≤ r) :
    |a * x| ≤ r := by
\end{lstlisting}
\sbsrow[2pt]
\textbf{Proof.}
{\[
  \begin{array}{@{}r@{\;}l@{\qquad}r@{}}
    |ax| &= |a|\,|x| & \text{(L4)}\\
         &\le 1\cdot|x| & \text{(L5--6)}\\
         &= |x| & \text{(L7)}\\
         &\le r & \text{(L8)}
  \end{array}
\]}
 After L4 the goal reads \lstinline!⊢ |a| * |x| ≤ r!; the \texttt{calc}
chain then proves the updated goal state: the first two steps use library lemmas
(\texttt{mul\_le\_mul\_of\_nonneg\_right}, \texttt{one\_mul}) and the last the hypothesis
(\texttt{hx}).
\sbsnext
\begin{lstlisting}[firstnumber=4]
  rw [abs_mul]   -- goal becomes  |a| * |x| ≤ r
  calc |a| * |x| ≤ 1 * |x| :=
        mul_le_mul_of_nonneg_right ha (abs_nonneg x)
    _ = |x| := one_mul |x|   -- 1 * |x| = |x|
    _ ≤ r := hx              -- the hypothesis hx (initial state)
\end{lstlisting}
\sbsrow[4pt]
\textbf{Claim 2.} For $x_{k+1}=ax_k$ with $|a|\le1$ and $|x_0|\le r$, $|x_k|\le r$ for
all $k$ (L14--16). The dynamics $x_{k+1}=ax_k$ from the initial state $x_0$ is
implemented by the \texttt{traj} function (L9--12).
\sbsnext
\begin{lstlisting}[firstnumber=9]
-- The trajectory x_k of x⁺ = a x, defined by recursion on k.
def traj (a x₀ : ℝ) : ℕ → ℝ
  | 0 => x₀                       -- x_0 = x₀
  | k + 1 => a * traj a x₀ k      -- x_{k+1} = a x_k

-- If |a| ≤ 1 and |x₀| ≤ r, then |x_k| ≤ r for every k.
theorem traj_bounded (a x₀ r : ℝ) (ha : |a| ≤ 1) (h0 : |x₀| ≤ r) :
    ∀ k, |traj a x₀ k| ≤ r := by
\end{lstlisting}
\sbsrow[2pt]
\textbf{Proof.} By mathematical induction on $k$ (L17--18).
\emph{Base} $k=0$: $x_0$ satisfies the bound by assumption (L19--21).
\sbsnext
\begin{lstlisting}[firstnumber=17]
  intro k              -- fix an arbitrary k : ℕ
  induction k with
  | zero =>            -- base case k = 0
    rw [traj]          -- unfold traj: goal becomes |x₀| ≤ r
    exact h0
\end{lstlisting}
\sbsrow[2pt]
\emph{Step}: assume $|x_k|\le r$ (the induction hypothesis \texttt{ih}). Then
$x_{k+1}=ax_k$ satisfies $|x_{k+1}|\le r$ by Claim~1 (L22--24). Note that a proved theorem is
used like a function: \texttt{interval\_invariant a r (traj a x₀ k) ha ih} (L24)
applies Claim~1 to $a$, $r$, $x_k$ and to the proofs \texttt{ha}, \texttt{ih} of its
hypotheses. In the step case, Lean
displays the goal state
\begin{lstlisting}[style=goal]
ih : |traj a x₀ k| ≤ r
⊢ |traj a x₀ (k + 1)| ≤ r
\end{lstlisting}
and after unfolding \texttt{traj} (L23) the goal becomes
\lstinline!|a * traj a x₀ k| ≤ r!, which is exactly Claim~1.
\sbsnext
\begin{lstlisting}[firstnumber=22]
  | succ k ih =>       -- step: from ih : |x_k| ≤ r prove |x_{k+1}| ≤ r
    rw [traj]          -- unfold traj: goal becomes |a * x_k| ≤ r
    exact interval_invariant a r (traj a x₀ k) ha ih   -- reuse Claim 1
\end{lstlisting}
\end{sbstab}
\end{sbsrows}

\section{Prediction: inverting the noise filter}\label{sec:inv}

An instructive use case for Lean is the analysis of existing results from the
literature. In this section we illustrate this on a classic lemma from prediction-error
identification, \citet[Lemma~3.1]{ljung1999sysid}, on the inversion of the noise
filter. We first report the lemma and its proof as given in the book. The Lean
formalization brings out a delicate point, the interchange of an infinite
double sum, which the proof takes for granted. Justifying it requires either a
stronger hypothesis or a weaker conclusion: for bounded noise the original
claim holds for every realisation, whereas for white noise with finite variance it holds instead almost surely,
but not for all realisations. We discuss both options, and state the resulting lemma certified in Lean.

\subsection{The original lemma}\label{sec:lemma}

Consider the disturbance
\begin{equation}\label{eq:disturbance}
v(t)=H(q)e(t)=\sum_{k=0}^{\infty}h(k)e(t-k),
\end{equation}
where the white noise $\{e(t)\}$ is a sequence of independent random variables with
zero mean and variance $\lambda$, $q$ denotes the forward time-shift operator,
$q e(t) = e(t+1)$, and $H(q) := \sum_{k=0}^{\infty}h(k)q^{-k}$ is stable,
\emph{i.e.}, $\sum_{k=0}^{\infty}|h(k)|<\infty$.

The noise model is \emph{invertible}~\citep[Sec.~3.2]{ljung1999sysid} if for known $v(s), s \leq t$, $e(t)$ can be computed as
\begin{equation}\label{eq:inv_noise}
    e(t)=\tilde H(q)v(t)=\sum_{k=0}^{\infty}\tilde h(k)v(t-k),
\end{equation}
with
$\sum_{k=0}^{\infty}|\tilde h(k)|<\infty$.
The following lemma characterizes the inverse noise filter $\tilde H(q)$ in terms of $H(q)$.
Lemma~\ref{lem:Ljung}, Remark~\ref{rem:stability} and the proof below are reproduced almost verbatim from \citet{ljung1999sysid}.

\begin{lemma}[{\citealp[Lemma~3.1]{ljung1999sysid}}]\label{lem:Ljung}
Consider $\{v(t)\}$ defined as in \eqref{eq:disturbance} and assume that $H$ is stable. Let
\begin{equation}\label{eq:filt}
H(z)=\sum_{k=0}^\infty h(k)z^{-k}
\end{equation}
 and assume that the function $1/H(z)$ is
analytic in $|z|\ge1$:
\begin{equation}\label{eq:invFilt}
\frac{1}{H(z)}=\sum_{k=0}^\infty\tilde h(k)z^{-k}
\end{equation}
Define the filter $H^{-1}(q)$ by $H^{-1}(q)=\sum_{k=0}^\infty\tilde h(k)q^{-k}$. Then, $\tilde H(q)=H^{-1}(q)$ satisfies \eqref{eq:inv_noise}.
\end{lemma}

\begin{remark}[\citealp{ljung1999sysid}]\label{rem:stability}
That \eqref{eq:invFilt} exists for $|z|\ge1$ also means that $H^{-1}(q)$ is stable:
$\sum_k|\tilde h(k)|<\infty$.
\end{remark}
\begin{proof}
\emph{Step A.} For $|z|\ge1$, multiplying \eqref{eq:filt} by \eqref{eq:invFilt} and collecting the powers by substituting $\ell=k+s$ we get,
\[
  1=\sum_{k=0}^{\infty}\sum_{s=0}^{\infty}h(k)\tilde h(s)z^{-(k+s)}
   =\sum_{\ell=0}^{\infty}\Bigl(\sum_{k=0}^{\ell}h(k)\tilde h(\ell-k)\Bigr)z^{-\ell},
\]
which implies that
\begin{equation}\label{eq:delta}
\sum_{k=0}^{\ell}h(k)\tilde h(\ell-k)=\delta(\ell) =\left\{\begin{matrix}
1 \ \rm{if} \ \ell=0 \\
0 \ \rm{if} \ \ell \neq 0
\end{matrix}\right.
\end{equation}
\emph{Step B.} Now let $\{v(t)\}$ be defined by \eqref{eq:disturbance} and consider:
\begin{align*}
\sum_{k=0}^{\infty}\tilde h(k)v(t-k)
&=\sum_{k=0}^{\infty}\sum_{s=0}^{\infty}\tilde h(k)h(s)e(t-k-s)\\
&=\sum_{\ell=0}^{\infty}\Bigl(\sum_{k=0}^{\ell}\tilde h(k)h(\ell-k)\Bigr)e(t-\ell)
  =e(t),
\end{align*}
where the second equality uses $\ell =k+s$ and the last equality uses \eqref{eq:delta}.
\end{proof}

\subsection{The interchange of the infinite double sum}\label{sec:swap}

In Step~B, the second equality swaps the summation order
of an \emph{infinite} double sum. The original sum adds the terms
$\tilde h(k)h(s)e(t-k-s)$ row by row (over $s$, then over $k$), but the rearranged sum is
along the antidiagonals $k+s=\ell$. Such a swap requires a justification; without
one, equality can fail, as Example~\ref{ex:swap} below illustrates.

\begin{example}[counterexample to Step~B]\label{ex:swap}
Let $H(q)=1-\tfrac12 q^{-1}$, which satisfies every hypothesis of
Lemma~\ref{lem:Ljung}, with $\tilde h(k)=2^{-k}$. For $e(t)=2^{-t}$,
$t\in\mathbb{Z}$, we get $v(t)\equiv0$. Hence
$H^{-1}(q)v(t)=\sum_k\tilde h(k)v(t-k)=0\neq e(t)$, whereas summing the same
terms along the antidiagonals, as in Step~B, gives $e(t)$.
\end{example}

Note that the counterexample involves a particular sequence $e$, which one would
hardly call noise. It grows into the past exactly as fast as $\tilde h$ decays, so
that the terms do not decay in $k$.
Such a sequence is however not excluded by Lemma~\ref{lem:Ljung}, hence the stated identity
does not hold for every possible realisation.
We remark that Lean did not discover the counterexample by itself. Rather, it \emph{refused} to verify the  interchange until a justification was supplied.

\subsection{Justifying the interchange}\label{sec:fixes}

We present two ways to justify the interchange in the summation: restrict the noise to a bounded
range, so that the original claim holds for every realisation, or allow
unbounded random noise and weaken the conclusion to hold almost surely.

\emph{Bounded noise.} By Fubini's theorem for series, the swap is valid if the
double series converges absolutely:
\begin{equation}\label{eq:AC}
  X_t:=\sum_{k,s\ge0}|\tilde h(k)|\,|h(s)|\,|e(t-k-s)|<\infty .
\end{equation}
If $e$ is bounded, $|e(t)|\le M$, then
$X_t\le M\|\tilde h\|_1\|h\|_1<\infty$, and the identity holds for every realisation.

\emph{Unbounded noise.} The bounded-noise hypothesis can be
too restrictive. If $e$ is assumed to be a white noise with finite variance $\lambda$ as in the book, its realisations may be unbounded. Nonetheless,
condition~\eqref{eq:AC} still holds \emph{almost surely}: by Tonelli's theorem,
$\mathbb{E} X_t\le \sup_s \mathbb{E}|e(s)|\,\|\tilde h\|_1\|h\|_1\le\sqrt\lambda\,
\|\tilde h\|_1\|h\|_1<\infty$, since $\mathbb{E}|e(s)|\le(\mathbb{E}e^2(s))^{1/2}=\sqrt\lambda$;
hence $X_t<\infty$ with probability one, for all $t\in\mathbb{Z}$. On each such
realisation Fubini's theorem applies, so the lemma
holds \emph{with probability one}, although not for each and every realisation.

\subsection{The precise statement}\label{sec:precise}

We now state the lemma in precise form for possibly unbounded random noise. The bounded case
follows from the same proof, with~\eqref{eq:AC} holding for every realisation.

\begin{lemma}[precise form]\label{lem:Ljung_new}
Consider $\{v(t)\}$, $H(z)$, $\tilde h(k)$ and $H^{-1}(q)$ as in
Lemma~\ref{lem:Ljung}, and assume:
\begin{enumerate}
  \item[(A1)] the filters $H$ and $H^{-1}$ are stable: $\sum_k|h(k)|<\infty$ and
    $\sum_k|\tilde h(k)|<\infty$;
  \item[(A2)] \eqref{eq:invFilt} holds, with a convergent series, for every $|z|\ge1$;
  \item[(A3)] $\sup_t\mathbb{E}|e(t)|<\infty$.
\end{enumerate}
Then, with probability~one, for every $t\in\mathbb{Z}$, the series defining $v(t)$
converges absolutely and
$H^{-1}(q)v(t)=e(t)$, that is,
$\tilde H(q)=H^{-1}(q)$ satisfies~\eqref{eq:inv_noise}.
\end{lemma}
(A1)--(A2) are the parts of Ljung's hypotheses on $H$ that the proof uses, and they
follow from those hypotheses when $1/H$ is analytic on a neighbourhood of $|z|\ge1$
(Remark~\ref{rem:stability}). (A3) holds for zero-mean white noise with finite
variance $\lambda$, since $\mathbb{E}|e(t)|\le\sqrt\lambda$.

The Lean proof of Lemma~\ref{lem:Ljung_new} is provided in our repository
(\leanlive{Lemma_3_1_Ljung}{\texttt{Lemma\_3\_1\_Ljung.lean}}).
It follows
Steps~A and~B of the proof of Lemma~\ref{lem:Ljung}; the only addition is that \eqref{eq:AC} is first
established almost surely (Section~\ref{sec:fixes}), which justifies the
interchange in Step~B.

In summary, because Lean accepts no step without justification, formalizing
Lemma~\ref{lem:Ljung} makes explicit that its conclusion holds
\emph{almost surely}, although not for every realisation. The
interchange of summations in Step~B requires a specific condition: the first-moment bound
(A3), which white noise with finite variance satisfies.
For bounded noise, the original claim holds for every realisation.

\section{Template Lean code for system identification}\label{sec:template}

This section presents a code template, so that readers can start formalizations of their own system identification results in Lean.

\emph{A four-step template}

\begin{enumerate}\setlength\itemsep{0pt}
\item \emph{Parameters and data.} Declare them as Lean types: a scalar parameter with type
  \lstinline!ℝ! (\emph{e.g.}, Box~\ref{box:toyid}), a parameter vector as a \texttt{structure} or
  \lstinline!Fin d → ℝ!, a signal as \lstinline!ℤ → ℝ!, and a random signal as
  \lstinline!ℤ → Ω → ℝ! on a probability space $(\Omega,\mu)$.
\item \emph{Model structure.} Define the model, or its predictor, with \texttt{def}, as a
  function of the parameter, \emph{e.g.}, \texttt{traj} in Box~\ref{box:scalar}, or the filter
  $H(q)e(t)$ of \eqref{eq:disturbance}, \texttt{filt} (see \texttt{Lemma\_3\_1\_Ljung.lean}).
\item \emph{Claim.} State the property as a \texttt{theorem} whose proof is stated as
  \texttt{sorry} initially, and every assumption is a named hypothesis, such as \texttt{hA1},
  \texttt{hA2}, \texttt{hA3} for (A1)--(A3) of Lemma~\ref{lem:Ljung_new}. Lean checks that
  the statement is well formed, and one can compare it, symbol by symbol, with the theoretical statement.
\item \emph{Proof.} Replace \texttt{sorry} one tactic at a time, reading the goal
  state; move intermediate claims into separate lemmas, such as Claim~1 of
  Box~\ref{box:scalar}, or  Steps~A and~B of the proof of Lemma~\ref{lem:Ljung}.
\end{enumerate}

 Three
definitions given in  \leanlive{Template}{\texttt{Template.lean}} can set the stage:
\begin{leanbox}[openlean=Template]
\begin{lstlisting}
-- a signal u(t), t ∈ ℤ
abbrev Signal := ℤ → ℝ
-- a model structure (Def. 4.3): M θ u y t is the prediction ŷ(t|θ)
abbrev ModelStructure (Θ : Type) := Θ → Signal → Signal → Signal
variable {Θ : Type}
-- global identifiability (Defs. 4.6-4.7)
def IsIdentifiable (M : ModelStructure Θ) : Prop :=
  ∀ θ₁ θ₂ : Θ, M θ₁ = M θ₂ → θ₁ = θ₂
\end{lstlisting}
\end{leanbox}
Here, $\text{ModelStructure}(\Theta) := \Theta \to (\mathbb{Z} \to \mathbb{R})_u \to (\mathbb{Z} \to \mathbb{R})_y \to (\mathbb{Z} \to \mathbb{R})_{\hat{y}}$ denotes the predictor in curried form: it takes the parameter $\theta$, the input $u$ and the output $y$, one at a time, and returns the predicted signal $\hat{y}(t|\theta)$.

\noindent The following listing applies the template to the ARX structure and proves
that it is strictly globally identifiable, as stated in \citet[Sec.~4.6]{ljung1999sysid}:
\begin{leanbox}[openlean=Template]
\begin{lstlisting}
-- Step 1: ARX model parameters. See Ljung's (4.8).
@[ext] structure ARXParams (na nb : ℕ) where
  a : Fin na → ℝ    -- a i is the coefficient of y(t-1-i)
  b : Fin nb → ℝ    -- b j is the coefficient of u(t-1-j)

-- Step 2: the model structure, given by its predictor (4.10)-(4.12).
def arx (na nb : ℕ) : ModelStructure (ARXParams na nb) :=
  fun θ u y t =>
    -∑ i, θ.a i * y (t - 1 - i.val) + ∑ j, θ.b j * u (t - 1 - j.val)

-- A unit pulse at time s: the test signal of the proof.
def pulse (s : ℤ) : Signal := fun t => if t = s then 1 else 0

-- Step 3: the claim, first stated with `sorry` as its proof.
theorem arx_isIdentifiable (na nb : ℕ) :
    IsIdentifiable (arx na nb) := by
  -- Step 4: the proof. A pulse in y (or u) isolates one coefficient.
  intro θ₁ θ₂ h                     -- h : arx θ₁ = arx θ₂
  have hM : ∀ u y t, arx na nb θ₁ u y t = arx na nb θ₂ u y t := by
    intro u y t; rw [h]             -- equal predictions everywhere
  ext i                             -- compare a i, then b i
  · -- u = 0, y = pulse at time -1-i: ŷ(0|θ) = -a i
    simpa [arx, pulse, Fin.val_inj] using hM 0 (pulse (-1 - i.val)) 0
  · -- u = pulse at time -1-i, y = 0: ŷ(0|θ) = b i
    simpa [arx, pulse, Fin.val_inj] using hM (pulse (-1 - i.val)) 0 0
\end{lstlisting}
\end{leanbox}

\section{Discussion on Lean proofs}\label{sec:discussion}
A potential concern is that checking a Lean proof may be more tedious than reviewing and understanding a conventional one.
In practice, however, step-by-step verification is handled entirely by Lean's kernel, while LLMs can assist in reading and writing the code.   Crucially, the kernel only checks proofs, not statements. A proof may be formally valid, yet verify a statement that differs from the intended theoretical claim. It is important to ensure that the formal statement in Lean, \emph{i.e.}, its assumptions and conclusion, faithfully represents the intended claim. Once the statement is verified, the correctness of the proof rests solely on the small Lean kernel.  The effort saved on line-by-line verification can then go to what matters most: judging whether the result is significant and its proof insightful.

In this paper, the Lean code was written with the help of the Claude Code AI assistant, and  the authors reviewed the statements and their match with the original results.

\section{Conclusions}

This paper presented a first perspective on the use of Lean in system
identification. Through a series of illustrative examples, we
showed how conventional mathematical arguments can be complemented by formal,
machine-checkable proofs. More generally, accompanying theoretical results with Lean proofs could
increase the reliability, transparency, and reproducibility of research.
Formal proofs may help authors detect missing assumptions, allow reviewers and
readers to verify results without manually checking every algebraic step, and
provide reusable libraries of certified results on which more advanced
contributions can be built.

LLMs can substantially reduce the effort required to produce
such formalizations by translating mathematical arguments into Lean code,
searching existing libraries, and proposing intermediate lemmas. Importantly,
regardless of whether a proof is written by a human or generated by an AI
system, every accepted proof is checked by Lean's kernel. Human supervision
nevertheless remains essential to ensure that the formal statement faithfully
represents the intended mathematical result and that its assumptions and
conclusions are scientifically meaningful.

We therefore envision Lean not as a replacement for conventional mathematical
reasoning, but as a companion to it. Pen-and-paper proofs provide intuition and
explanation, while formal proofs provide an additional layer of and
reproducible verification. Developing this combination further may contribute
to stronger and more verifiable theoretical foundations for system
identification and control.

\section*{Acknowledgments}
The activities of Marco Forgione have been supported by the HASLER STIFTUNG under the project ARCADE: Agentic Research for Controller Architecture Discovery and Evaluation.

\bibliographystyle{plainnat}
\bibliography{references}

\end{document}